\documentstyle[preprint,eqsecnum,aps]{revtex}
\begin{document}
\draft
\tighten
\title{A point mass in an isotropic universe. Existence, uniqueness
and basic properties.}
\author{Brien C. Nolan\footnote{e-mail: nolanb@ccmail.dcu.ie}\\
School of Mathematical Sciences,\\
Dublin City University,\\ Glasnevin, Dublin 9,\\
Ireland.}
\maketitle
\begin{abstract}
Criteria which a space-time must satisfy to represent a point
mass embedded in an open Robertson--Walker (RW) universe are given. It is shown that McVittie's
solution in the case $k=0$ satisfies these criteria, but does not in the
case $k=-1$. Existence of a solution for the case $k=-1$ is proven and
its representation in terms of an elliptic integral is given. The following 
properties of this and McVittie's $k=0$ solution are studied; uniqueness,
behaviour at future null infinity, recovery of the RW and Schwarzschild limits,
compliance with energy conditions and the occurence of singularities.
Existence of solutions representing more general spherical objects embedded
in a RW universe is also proven.
\newline
\newline
\end{abstract}


\newcommand\newc{\newcommand}
\newc{\mcv}{McVittie}
\newc{\sch}{Schwarzschild\,}
\newc{\be}{\begin{eqnarray}}
\newc{\ee}{\end{eqnarray}}
\newc{\mass}{\tilde{M_2}}
\newc{\nn}{\nonumber}
\newc{\scri}{{\cal{I}}}
\newc{\mb}{\bar{m}}
\newc\mud{{\mu^\prime}}
\newc\mdd{\mu^{\prime\prime}}
\newc\mudd{\mu^{\prime\prime}}
\newc\nudd{\nu^{\prime\prime}}
\newc\nud{\nu^\prime}
\newc\ndd{\nu^{\prime\prime}}
\newc\mut{\dot{\mu}}
\newc\mutt{\ddot{\mu}}
\newc\mudt{\dot{\mud}}
\newc\nut{\dot{\nu}}
\newc\bedot{{\dot{\beta}}}
\newc\btt{{\ddot\beta}}
\newc\hd{h^\prime}
\newc\hdd{h^{\prime\prime}}
\newcommand{\rin}{{r\rightarrow\infty}}
\newcommand{\xin}{{x\rightarrow\infty}}
\newcommand{\dom}{{\cal{D}}}
\newcommand{\norm}[1]{|\!|{#1}|\!|_{\dom}}



\section{Introduction}
This paper deals with the embedding of massive objects in 
Robertson--Walker (RW) universes.
There are three ways in which the physical embedding may be modelled and thus 
treated mathematically. Firstly, one can treat the body as a test body 
whose dynamics are described by a suitable set of equations of motion (e.g. 
geodesic equations for a test particle, equations derived from the Nambu action
for cosmic strings). Secondly, the history of the surface of the object can be treated
as a boundary $\Sigma^-$, which is then matched with a diffeomorphic surface
$\Sigma^+$ in the `exterior' RW geometry. The usual matching conditions
are continuity of the first and second fundamental forms of $\Sigma\cong\Sigma^\pm$
\cite{israel}.
This technique has been used to study the formation and evolution of voids in cosmology
\cite{sato}, and to study domain walls \cite{vilenkin}.

The third method is to solve Einstein's field equations, exactly or approximately,
in such a way that the resulting solution can be interpreted as an embedding of some
massive object in a RW background. 
Two landmark papers in this vein are those of McVittie \cite{mcv},
who gave solutions of Einstein's equations with perfect fluid source 
which have been claimed to represent
the embedding of the Schwarzschild\, field in the three families $(k=0,\pm1)$ of RW space-times, and of Hawking 
\cite{hawk}, who studied gravitational radiation from a bound source in the 
$k=-1$ dust filled RW space-time. 
There also exist several papers dealing with
the {\em superposition} of the Kerr--Newman and RW space-times (see \cite{othersols},
or \cite{kraz} for a summary).
We stress that we are considering
the embedding of extended objects of finite size, so that the extensive studies
(see e.g. \cite{ellis-bruni}) of perturbations which occur throughout the universe
do not concern us here.

These three approaches incorporate various degrees of coupling
between the mass--energy of the extended body and the geometry of the universe at
large. In particular physical situations, one of the three provides an
appropriate model. For example, in cosmology, the galactic source of some observable
effect is treated as a test body moving on a time-like geodesic of the
RW geometry; the Einstein--Straus vacuole \cite{e-s} provides a
description of the effect of the cosmic expansion on the gravitational
field of the sun, but McVittie's solution is a more appropriate description
of the gravitational field outside a super-massive spherical body in an
otherwise uniform RW space-time. 

An interesting result was derived recently by Senovilla and Vera regarding the
cylindrical analogue of these models \cite{seno}. String dynamics in an 
RW universe deals with extended bodies which are limits  of cylindrical
objects and is well understood. However, moving to the next level of
coupling, the aforementioned authors showed that no static cylindrical region
can be matched continuously to a RW universe. 
Similiar results have also been obtained for the axially symmetric case
\cite{mars}.
Since real strings have internal structure,
this implies that at this level, strings cannot be embedded in a RW
universe. The same is true for any static locally cylindrical objects; coins,
bottles and (true) cylinders.
This begs the question: Can the third type of embedding be carried out for cylindrical objects? 
i.e. can we find an exact solution of Einstein's equation representing a
cylindrical object embedded in a RW space-time?

We will not attempt to answer this question here, but by examining carefully
the spherical case, suggest how the problem may be approached. Thus we readdress
the problem first discussed by McVittie, but from a modern point of view.

We find that McVittie's solution in the case $k=0$ satisfactorily describes
a massive particle embedded in a RW universe, but that his $k=-1$
solution does not. Motivated by the differences between these two solutions,
we lay down {\em a priori} conditions that a space-time $(V,g)$ must satisfy
to represent a massive particle embedded in a RW space-time.
We provide a solution in the case $k=-1$ and discuss uniqueness
in each case, which has not been done before. 
We emphasise that we have not found a new solution of Einstein's equations,
but have determined the physical significance of a certain class of
shear-free spherically symmetric perfect fluid solutions (see chapter 14 of
\cite{ksmh}).
Furthermore, we discuss the following
properties (mathematical and physical) of the solutions: (a) representation of
the $k=-1$ solution by an elliptic integral, (b) recovery of the Schwarzschild solution
in the vacuum limit, (c) behaviour at future null infinity, (d) compliance with 
energy conditions and (e) existence and nature of the central singularity.
A central tool in this analysis is Hawking's quasi-local mass \cite{hawk}. We show 
by this example how such quasi-local constructions can be used to obtain boundary
conditions for Einstein's equations useful for obtaining solutions in particular
situations. Since we are dealing with asymptotics in open space-times, the
case $k=+1$ is excluded from our discussion.

The structure of the paper is as follows.
In the next section we review McVittie's solution and the Hawking mass and 
point out problems with the interpretation of the former in the case $k=-1$.
In \S III, we set out conditions for a space-time $(V,g)$ to represent a 
massive particle embedded in a RW universe. Using these conditions and
Einstein's field equations, we show how the problem reduces to finding a
solution, with a certain asymptotic behaviour,
 of a non-linear second order differential equation. In \S IV, we prove the
existence of such a solution, and discuss uniqueness. Using this solution,
the properties listed above are discussed in \S V, and we make some concluding
comments in \S VI.
The global struture of these space-times is to be analysed in a subsequent paper.

\section{McVittie's Solution and the Hawking Mass}
In 1933, McVittie\, \cite{mcv} found solutions of Einstein's field equations for
a perfect fluid energy-momentum tensor, representing a Schwarzschild field
embedded in the RW space-times. 
His solutions can be written\cite{hog1} (using units in which $c=G=1$)

\begin{eqnarray} ds^2&=&-\left({1-m/2w\over1+m/2w}\right)^2 dt^2 +
e^\beta\left(1+{m\over 2w}\right)^4\left\{dr^2 +h^2(d\theta^2+\sin^2
\theta d\phi^2)\right\}\,, \label{mcv}\ee

where
\begin{eqnarray} m&=&m(t)\,,\qquad \beta\,\,=\,\,\beta(t)\,,\qquad 
\dot{\beta}\,\,=\,\,-2(\dot{m}/m)\,,\label{mass1}\ee
and here and throughout, a dot indicates partial differentiation
with respect to $t$ (a prime will be used for differentiation w.r.t. the
variable $r$). 
The functions $h(r), w(r)$ depend on a choice of 
$k(=-1,0,+1)$, the Riemannian curvature of the surfaces of homogeneity 
$t=$const. in the background RW universe;
\[ h(r)=
\left\{
\begin{array}{cl}
\sinh r,&k=-1;\\
r,&k=0;\\
\sin r,&k=+1;
\end{array}\right. \,\,
 w(r) =
\left\{
\begin{array}{cl}
2\sinh \frac{r}{2},&k=-1;\\
r,&k=0;\\
2\sin \frac{r}{2},&k=+1.
\end{array}\right.\]

The isotropic pressure $p_{\rm mv}$
and the energy density $\rho_{\rm mv}$ obtained from Einstein's field equations
are given by
\begin{eqnarray} 8\pi p_{\rm mv} &=& -{3\over 4}\bedot^2-
\btt\left({1+{m\over 2w}} \over {1-{m\over 2w}} \right)-
{ k e^{-\beta} \over
\left(1-{m\over 2w}\right)\left(1+{m\over 2w}\right)^5 } \,,
\label{mcvp}\\
8\pi \rho_{\rm mv} &=& {3\over 4}\bedot^2+{3k e^{-\beta}\over (1+{m\over 2w})^5 }\,.
\label{mcvr}\ee
The properties of this solution have been summarized by Raychaudhuri as follows 
({\em cf.}\cite{ray}, p.97):
``The McVittie\, solution follows uniquely under the following
conditions: (i) The line element is spherically symmetric with a singularity
at the centre. (ii) The energy--stress tensor is that of a perfect fluid. 
(iii) The fluid motion is shear free. (iv) The metric must asymptotically
go over to the isotropic cosmological form.'' (It should be noted that neither a proof
of this statement, nor a reference to one is given; McVittie's {\em ad hoc} approach does not include such a proof.)

The function $m=m_0e^{-\beta/2}$ for some constant $m_0$ by (\ref{mass1}), and is interpreted
as the mass at the singularity. When this is set equal to zero, the line element 
(\ref{mcv}) is that of a RW space-time.

The characterization of McVittie's solution quoted above is unsatisfactory, as points
(i) and (iv) refer to properties which are deduced simply by looking at the metric
tensor components relative to the line element (\ref{mcv}). We can show that point (iv) in 
particular is misleading. This point seems to imply that the solution 
corresponds to a point mass embedded in the RW geometry, so that the 
gravitational field is asymptotically that of a RW space-time.

Hawking \cite{hawk} has made this notion precise with a renormalized
(against the RW background) quasi--local mass measured at future null infinity $\scri^+$.
Since we are dealing with asymptotic
regions of the space--time, we restrict our attention to the
cases $k=-1,0$.

The Hawking mass is defined by analogy with the Bondi mass \cite{bondi} of a
bound source of gravitation in an asymptotically flat space--time; it
measures the mass of a bound source of gravitation in an
asymptotically RW universe. The additions to the total (infinite) mass
from the RW background are subtracted away in a gauge invariant
manner, as we describe now. The construction is valid
in any space--time.

We use the null tetrad $\{l^a,n^a,m^a,\bar{m}^a\}$, where
$l^a$ is chosen to be an outgoing null vector, and take $v$ to be an
affine parameter along the integral curves of $l^a$, so that 
\begin{eqnarray} l^a&=&{dx^a\over dv}\,.\label{aff}\ee
Taking $S$ to be a space--like 2--sphere orthogonal to $l^a$ and $n^a$
(so that $v=$constant on $S$), 
the quasi--local mass surrounded by $S$ is defined to be 
\begin{eqnarray} M(S)&=&\kappa \int (-\Psi_2-\sigma\lambda+\Phi_{11}+\Lambda) \,dS\,,\ee
where
\[ \kappa={1\over (4\pi)^{3/2}} \left(\int dS\right)^{1/2}\]
and the terms in the integrand have their usual meanings in 
Newman--Penrose notation. In the appropriate limits, $M$ yields the Bondi mass and
the ADM mass, and is the spherical version of Hayward's improved
quasi-local mass \cite{haymass}.

The renormalization is carried out by subtracting the local (fluid) matter which manifests itself in the
Ricci tensor terms, and to leave the non-local gravitational terms. 
To do this, Hawking \cite{hawk} has defined 
\begin{eqnarray} M_1(S)&=&\kappa \int (\Phi_{11}+\Lambda) \,dS\,,\label{hmass1}\ee
and
\begin{eqnarray} M_2(S)&=&\kappa \int (-\Psi_2-\sigma\lambda) \,dS\,.\label{hmass2}\ee

In order that the bound source mass is measured at $\scri^+$, 
\begin{eqnarray} \mass&=& \lim_{v\rightarrow \infty} M_2\,,\label{hmass} \ee
is defined to be the mass of the model.

Using a suitable null-tetrad, we can evaluate
(\ref{hmass2}) for McVittie's space-time (notice that due to spherical symmetry $\lambda=\sigma=0$).
We find
\begin{eqnarray} M_2(S)&=& m_0{h^5\over w^5}\,.\label{masval}\ee
Thus for $k=0$, wherein $h=w=r$, we have $M_2=m_0$, 
and so the Hawking mass is
\begin{eqnarray} \mass&=&m_0\,,\label{msk=0}\ee
which verifies the interpretation of $m_0$ as being 
the mass of a point particle embedded in the RW cosmos in this case.
However in the case $k=-1$, since $r\rightarrow\infty$ as $v\rightarrow\infty$,
we find that
\begin{eqnarray} M_2(S)&\rightarrow&\infty\,,\qquad{\rm as}\qquad v\rightarrow\infty\ee
and so the renormalized mass is infinite.
Thus the $k=-1$ McVittie solution {\em does not} represent a point mass embedded in a 
RW space-time.

The main aim of this paper, then, is to provide a solution of Einstein's equation which
does represent a point mass embedded in the $k=-1$ RW space-time.

\section{Description of Space-Time representing a point mass embedded in an
RW universe}

In this section, we give three conditions {\bf(C1-C3)} on space-time $(V,g)$ which,
if these conditions are satisfied, we postulate to represent a point mass
embedded in a RW universe. These conditions are motivated by the discussion 
above. For convenience, we will use the symbol $(M,g)$ to refer to such a space-time.

{\it {\bf Condition C1.} $(M,g)$ is spherically symmetric with a shear-free perfect fluid 
energy-momentum tensor.}

Demanding a perfect fluid energy-momentum tensor and spherical symmetry are obvious 
requirements for the space-time we seek to describe; the requirement that the
fluid flow lines be shear-free is a convenience that eases integration of the
field equations without ruling out the existence of a solution. There does not
seem to be any {\it a priori} reason why the shear should be set equal to zero.

Under these assumptions, the line element can be written as \cite{ksmh}
\begin{eqnarray} ds^2 &=& -e^\nu dt^2+e^\mu\left\{dr^2+h^2(r)d\omega^2\right\}\,,
\label{1}\ee
where $\nu=\nu(r,t)$, $\mu=\mu(r,t)$ and $h$ is an arbitrary function of $r$, which we may always assume is one of the functions $h(r)$ of \S II. 
Doing so maintains the connection with the corresponding forms of the RW space-times.

The density and pressure obtained from Einstein's field equations are
given by
\begin{eqnarray} 8\pi \rho &=& {3\over 4}\mut^2 e^{-\nu} - e^{-\mu} \{\mdd + {1\over 4}
{\mud}^2 + 2{\hd \over h} \mud + 3 {\hdd \over h} \} \,,
\label{2}\\
8\pi p&=& e^{-\mu} \{ {1\over 2}(\mdd + \ndd)+
{1\over 4}{\nud}^2 +  {1\over 2}{\hd \over h}
(\nud + \mud) + {\hdd \over h} \} - e^{-\nu} \{\mutt + {3\over 4}\mut ^2-
{\mut \nut \over 2} \} \,.
\label{3}\ee

The remaining field equations reduce to \cite{ksmh}
\begin{eqnarray} e^\nu&=& \mut^2e^{-g(t)}\,,\label{4}\\
\mudd-\frac{1}{2}\mud^2-\frac{h^\prime}{h}\mud&=&F(r)e^{-\mu/2}\,,\label{5}
\ee
where $g(t)$ and $F(r)$ are arbitrary functions of their arguments.

The Weyl tensor is Petrov type $D$, and on a naturally occurring null tetrad,
the only non-zero Newman-Penrose component is
\begin{eqnarray} \Psi_2&=&-\frac{1}{6}F(r)e^{-3\mu/2}\,.\label{6}\ee
Then the Hawking mass enclosed by any $t=$ constant $r=$ constant surface $S$
is
\begin{eqnarray} M_2(S)&=&\frac{1}{6}h^3(r)F(r)\,.\label{7}\ee
This surface $S$ is a metric sphere of radius $R(r,t):=h(r)e^{\mu/2}$. In the solutions we
examine, $\partial R/\partial r>0$, and so since $M_2(S)$ is independent of $t$,
the limit $r\rightarrow\infty$ 
of (\ref{7}) yields the Hawking mass at infinity.

The second condition achieves two things. Firstly, it identifies the RW background
against which the Hawking mass is measured, and secondly, it ensures that the limit
$r\rightarrow\infty$ has physical significance. This condition is a minimal requirement
that $(M,g)$ `looks like' a RW space-time near infinity.

{\it {\bf Condition C2.}
\[ \lim_{r\rightarrow\infty} \mu(r,t) =\beta(t)\]
for all $t$ in the range of $\mu$, and where $\beta$ is the function appearing
in the line element of the RW universe,
\begin{eqnarray} ds^2&=& -dt^2 + e^{\beta(t)}\left\{dr^2+h^2(r)d\omega^2\right\}\,.
\label{8}\ee}

As we shall see below, it is not necessary to give the corresponding condition
for $\nu$ ($\nu\rightarrow 0$ as $r\rightarrow\infty$), as this 
is ensured by a naturally arising choice of function of integration.

We turn now to the Hawking mass and consider how this may be used in our description.
We wish to embed a finite non-zero mass in the RW universe, indicating that $M_2(S)$
should yield a finite positive number when measured at infinity, i.e.
\begin{eqnarray} \lim_{r\rightarrow\infty} M_2(S)&=&m_0 \label{9}\ee
for some positive constant $m_0$.
However this allows the possibility of surplus Hawking mass arbitrarily close
to infinity. We will therefore consider the stronger condition,
\begin{eqnarray} M_2(S)&=&m_0\,,\label{10}\ee
for some positive constant $m_0$.
This condition suggests that there is an isolated body of mass $m_0$ at the centre
of the space-time. In order to present a more general discussion in \S IV, we will deal 
with the condition (\ref{9}), but focus on the special case of (\ref{10}).

We see from the above that (\ref{9}) implies
\[ F(r)=O(h^{-3})\qquad {\rm as}\quad r\rightarrow\infty\,,\]
while (\ref{10}) gives
\begin{eqnarray} F(r) &=& \frac{6m_0}{h^3(r)}\,.\label{12}\ee

This last equation is interesting, as it is a necessary and sufficient condition
for the energy density to be spatially homogeneous, i.e.
\begin{eqnarray} F(r)={\rm constant}\times h^{-3}(r) \iff \rho=\rho(t)\,.\label{13}\ee

Thus to obtain the constant Hawking mass condition (\ref{12}), or equivalently to
express the fact that there is no `extra' energy density in the universe outside
the embedded mass, we take the third condition to be as follows.

{\it {\bf Condition C3.} The energy density (\ref{2}) obeys $\rho=\rho_0(t)$, where
$\rho_0(t)$ is the energy density calculated via Einstein's field equations
of the RW space-time with line element (\ref{8}). Furthermore, the constant
quantity $m_0:=h^3(r)F(r)/6$ is positive.}

We can show now how to obtain $\nu\rightarrow\infty$
as $\rin$.
From the above, the energy-density of $(M,g)$ is given by
\[ 8\pi\rho=\frac{3}{4}e^{g(t)}+3ke^{-\beta(t)}\,,\]
while that of the RW background is given by
\[ 8\pi\rho_0=\frac{3}{4}{\dot{\beta}}^2 + 3ke^{-\beta(t)}\,.\]
($\rho_0$ and $p_0$, the energy-density and pressure of the RW
background may be read off from (\ref{mcvr}) and (\ref{mcvp}) respectively
by taking $m=0$.)
Thus $\rho(t)=\rho_0(t)$ requires $e^{g(t)}={\dot{\beta}}^2$.
We will see below that there exists a solution 
satisfying the three conditions above for which $\mu$ is differentiable with respect to $t$,
and which obeys $\lim_\rin \mut ={\dot{\beta}}.$ Thus
\[ \lim_{\rin} e^\nu =\lim_\rin \mut^2 e^{-g(t)} =1\,,\]
as claimed.

The physical problem has been modelled using the conditions {\bf C1-C3} above.
We see that the remaining mathematical analysis is to find a solution $\mu(r,t)$
of
\begin{mathletters}\label{14a}
\begin{eqnarray} e^{\mu/2}(\mu^{\prime\prime}-\frac{1}{2}{\mu^\prime}^2-\frac{h^\prime}{h}
\mu^\prime) &=& \frac{6m_0}{h^3}\,,\label{14}\ee
with the boundary condition
\begin{eqnarray} \lim_{r\rightarrow\infty} \mu(r,t) &=&\beta(t)\,,\label{15}\ee
\end{mathletters}
for all $t$ in the range of $\mu$.

More generally, we look for a solution of (\ref{14}) with the right hand side replaced by 
$O(h^{-3}(r))$ as $r\rightarrow\infty$. Our analysis below is based on this version
of the equation, which corresponds to the condition (\ref{9}).

We note at this stage that McVittie's solution \cite{mcv} in the case $k=0$
satisfies conditions {\bf C1-C3}. We focus henceforth on the case $k=-1$,
and work under this assumption. 

We conclude this section by giving a useful transformation which will make the
problem (\ref{14a}) more manageable.
Defining $\gamma=e^{-\mu/2}$ and $x=w^2$ (c.f. \S II), these become respectively
\begin{mathletters}\label{16}
\begin{eqnarray} \gamma_{xx} &=&G(x)\gamma^2\,,\label{16a}\\
\lim_\xin\gamma&=&e^{-\beta/2}\,,\label{16b}
\ee\end{mathletters}
where
\[ G(x) = -24m_0(x^2+4x)^{-5/2} = O(x^{-5}) \qquad {\rm as}\quad \xin \,,\]
and the subscript indicates partial differentiation with respect to $x$.
Using $G(x)=O(x^{-5})$ corresponds to the general case (\ref{9}).
We now proceed to prove existence of a solution of the boundary value problem
(\ref{16}).

\section{Existence of a Solution and Uniqueness Considerations}
Writing $\gamma(x,t)=a(t)(1+Y(x,t))$ where $a(t):=e^{-\beta(t)/2}$ allows us
to restate the problem (\ref{16}) as
\begin{eqnarray} Y_{xx}&=&aG+2aGY+aGY^2\,,\nn\\
Y(x,t)&=&o(1)\,,\qquad\xin\,.\nn
\ee
We treat $x$ as a complex variable and solve the equation in a neighbourhood
of infinity which includes $\{x\in{\cal {C}}:\Im{x}=0, \Re{x}>x_0>0\}$ where 
$x_0$ is some real constant.

Consider the equation
\begin{eqnarray} Y_{xx}&=&aG+\delta(2aGY+aGY^2)\,,\label{17}\ee
where $\delta>0$ is a small parameter.
We look for a solution of this equation of the form
\begin{eqnarray} Y(x,t)&=&\sum_{n=0}^\infty\delta^nY_n(x,t)\,,\label{18}\ee
and having found one, show that for sufficiently large values of $x$, this
converges in the limit $\delta=1$ and obeys $Y=o(1)$ as $\xin$. This will prove
existence of the required solution. We will use the ansatz $Y(x,t)=a^{n+1}(t)y_n(x)$,
and throughout this section, a prime on $y_n$ indicates differentiation
with respect to argument.

To proceed, we fill out (\ref{17}) using (\ref{18}), and equate powers of $\delta$.
This leads to the system of equations
\begin{mathletters}\label{19}
\begin{eqnarray} y_0^{\prime\prime}&=&G(x)\,,\label{19a}\\
y_n^{\prime\prime}&=&2G(x)y_{n-1}+G(x)\sum_{m=0}^{n-1}y_my_{n-1-m}\,,\quad n\geq1\,.
\label{19b}\ee
\end{mathletters}
Clearly, this system admits solutions obeying
\[ y_n(x)=o(1)\,,\qquad y_n^\prime(x)=o(x^{-1})\,,\qquad \xin\,.\]
Then we can write
\begin{eqnarray*}
 y_n(x) &=&\int_\infty^x y_n^\prime(s)\,ds\\
&=&\int_\infty^x (x-s)y_n^{\prime\prime}(s)\,ds\,.
\end{eqnarray*}
Transforming the integral to one over a finite contour via $z=s^{-1}$ and using 
basic bounds for integrals, we obtain
\[ |y_n(x)|\leq 2|x^{-1}| \sup_{\{|x|<|s|\}} |s^3y_n^{\prime\prime}(s)|\,.\]
Then taking $\dom$ to be a neighbourhood of infinity contained in the intersection
of $\{x\in{\cal C}:|x|>x_0>0\}$ and a sector containing the positive real axis, we have
\[ |y_n(x)| \leq C\norm{W_n}\]
for all $x\in\dom$, where $C:=2x_0^{-1}$, $W_n(x):=x^3y_n^{\prime\prime}(x)$ and
$\norm{\cdot}$ is the supremum norm restricted to $\dom$. So
\begin{eqnarray} \norm{y_n}\leq C\norm{W_n}\,.\label{20}\ee
Applying these definitions to (\ref{19b}), we obtain
\[\norm{W_n}\leq 2B\norm{y_{n-1}}+B\sum_{m=0}^{n-1} \norm{y_m}\norm{y_{n-1-m}}\,,\]
where $B=B(x_0)=\norm{x^3G(x)}$, and so using (\ref{20}), 
\[\norm{y_n}\leq 2A\norm{y_{n-1}}+A\sum_{m=0}^{n-1} \norm{y_m}\norm{y_{n-1-m}}\,,\]
with $A=BC$.

Using this inequality, we can derive a geometric bound for the $\norm{y_n}$,
which will suffice to prove the convergence properties of $\sum y_n$ required to
show that this {\it formal} solution is a (convergent) solution.
To see how, define the sequence of positive reals $\{b_n\}_{n=0}^{\infty}$ by
\begin{eqnarray} b_0&=& \norm{y_0}\,,\nn\\
b_n&=&2b_{n-1}+\sum_{m=0}^{n-1}b_mb_{n-1-m}\,.\nn\ee
Then we see from the last inequality that
\begin{eqnarray} \norm{y_n}\leq A^n b_n\,, n\geq0\,.\label{ineq}\ee

Consider the formal power series
\[ P(X):=\sum_{n=0}^\infty b_nX^n\,.\]
From the recurrence relation for the $b_n$, we find that $P(X)$ obeys
\[ P = b_0 +X(2P+P^2)\,,\]
the solution of which consistent with the definition of $P$ is
\begin{eqnarray} P(X) &=&\frac {1-2X- ((1-2X)^2 -4b_0X)^{1/2}}{2X}\,.\label{21}\ee
This is an analytic function of $X$ in a neighbourhood of the origin, and
so for any $0<\lambda\in{\cal{R}}$ with $|\lambda|<$ radius of convergence of $P(X)$,
\[ \sum_{n=0}^\infty b_n\lambda^n\]
is convergent. Then each term in this series must be bounded, i.e. there exists
some positive real constant $K$ such that
\[ b_n\lambda^n < K\,,\qquad n\geq 1\,,\]
and so 
\[ \norm{y_n} \leq K\left(\frac{A}{\lambda}\right)^n\,,\qquad n\geq1\,.\]
Then for $x_0$ sufficiently large, this last inequality will read
\begin{eqnarray} \norm{y_n}\leq \kappa^n\,,\label{22}\ee
for some $0<\kappa<1$.
To see this, notice that as $x_0$ increases, the region $\dom$ gets smaller, so that
$A=2x_0^{-1}\norm{x^3G(x)}$ decreases (recall that $G(x)=O(x^{-5})$). $b_0=\norm{y_n}$
is non-increasing, which by (\ref{21}) indicates that the radius of convergence of 
$P(X)$ in non-decreasing, allowing the use of non-decreasing values of $\lambda$.

The condition (\ref{22}) is sufficient to imply that
\[ \sum_{n=0}^\infty y_n(x)\]
converges uniformly on $\dom$ (see e.g. \cite{remmert}).
Hence
\[ Y(x,t) =\sum_{n=0}^\infty a^{n+1}(t)y_n(x)\]
converges uniformly on some subset of $\dom\times{\cal{R}}$ $(x\in\dom,t\in{\cal{R}})$.
By our construction, this is a solution of (\ref{17}) in the case $\delta=1$ obeying
$Y(x,t)=o(1)$ as $\xin$.

Furthermore, for each fixed value of $x$, the series
\[ \sum y_n(x)a^{n+1}(t)\,,\quad \sum y_n(x)(n+1)a^n(t)\frac{da}{dt}\]
are both uniformly convergent on some interval of the real $t$-axis (which will
contain the set $\{t\in{\cal{R}}:|a(t)|<1\}$). Hence by standard results
\cite{apostol}, $\partial Y/\partial t$ exists on this interval, and
\[ \frac{\partial Y}{\partial t} = \sum_{n=0}^\infty \frac{\partial Y_n}{\partial t}(x,t)\,.\]

We summarise and extend as follows.
\newline
{\it
{\bf Theorem 1 (Existence and uniqueness)}
There exists $0<x_0\in{\cal{R}}$ and a non-empty subset $A\subseteq{\cal{R}}$
such that on $\{x\in{\cal{R}}:|x|>x_0\}\times A$, there exists a solution of
$\gamma_{xx}=G(x)\gamma^2$ where $G(x)=O(x^{-5})$ as $\xin$, obeying
$\gamma(x,t)=e^{-\beta(t)/2}+o(1)$ as $\xin$. This solution is differentiable
with respect to $t$. Furthermore, if $G(x)$ is analytic in a neighbourhood
of infinity, then this solution is the unique analytic solution.}

For the proof of the last statement, note first of all that if $G(x)$ is analytic in a neighbourhood of infinity, 
then the 
existence proof of the theorem gives the construction of an analytic
solution of (\ref{16}). For each term in the series (\ref{18}) is found by integrating
analytic functions, and is therefore analytic; uniform convergence guarantees
analyticity of the sum.
Next, we argue that if $G(x)$ and a solution $\gamma(x,t)$ of (\ref{16})
are analytic functions
of $x$ in a neighbourhood of infinity, then this solution is unique.
For let $\gamma_1, \gamma_2$ be two such solutions, and define
\[ \Gamma:=\gamma_1-\gamma_2\,,\quad H(x,t):=(\gamma_1+\gamma_2)G(x)=O(x^{-5})\,.\]
Then $H$ is analytic in $x$ in a neighbourhood of infinity, and $\Gamma$ obeys
\begin{mathletters}\label{23}
\begin{eqnarray} \Gamma_{xx}&=&H(x,t)\Gamma\,,\label{23a}\\
\Gamma&=&o(1)\,,\xin\,.\label{23b}\ee
\end{mathletters}
By analyticity, the solutions of this linear differential equation, which has 
a regular singular point at infinity, can be written in the form
\[ \Gamma = \sum_{n=0}^\infty a_n(t)x^{-n+p}\]
for some real $p$. Using a Taylor series expansion about $x^{-1}=0$ for $H(x,\cdot)$
and filling out the equation (\ref{23a}), we find that  the two independent
solutions are described by
\newline
(i) $p=0$, $a_0$ arbitrary, $a_1=a_2=0$, $a_n, n\geq3$ determined by recurrence
relations, and proportional to $a_0$;
\newline
(ii) $p=1$, $a_0, a_1$ arbitrary, $a_2=0$, $a_n, n\geq3$ determined by 
recurrence relations given by linear combinations of $a_0$ and $a_1$.

In either case, we see that non-zero solutions do not obey $\Gamma=o(1)$ as
$\xin$, so that $\Gamma\equiv 0$ is the only solution of (\ref{23}), and so
$\gamma_1=\gamma_2$, proving uniqueness.

In particular, in the case of most interest to us, $G(x)=-24m_0(x^2+4x)^{-5/2}$
is such a function, and so the solution produced by the theorem is the unique analytic solution.
Notice also that when $m_0=0$ in this solution, we obtain the RW line
element (\ref{8}).
It will be useful to have the first few terms of this solution.
These terms are obtained by integrating (\ref{19a}) with the appropriate choice of integration
constants, and yield
\begin{eqnarray} \gamma(x,t) &=& e^{-\beta/2}\{1-m
(\frac{2}{(x^2+4x)^{1/2}} -(x+2) +(x^2+4x)^{1/2})\}
 + O(x^{-6})\,,\label{24}\ee
where $m=m_0e^{-\beta/2}$.
Converting to the original coordinates using the transformation given
prior to (\ref{16}), this leads to
\begin{eqnarray} e^\mu &=&e^{\beta(t)}(1-4m_0e^{-\beta/2}e^{-3r}) + O(e^{-5r})\,.
\label{25}\ee

We note that the corresponding first order term in McVittie's $k=-1$ solution
is
\[ e^\mu = e^\beta(1+2m_0e^{-\beta/2}e^{-r/2}) + O(e^{-3r/2})\,,\]
from which we can identify the problem with this solution; the metric coefficients
do not tend to those of the RW metric rapidly enough.

We now turn our attention briefly to the case $k=0$. 
The existence proof of this section is of course not needed for this case,
because as we have seen already, McVittie's $k=0$ solution satisfies the
conditions {\bf C1-C3}.
The uniqueness result of this section {\it does} apply, and so we see
that if Raychauduri's conditions quoted in \S II are replaced by the conditions
of \S III, then under the added hypothesis of analyticity, the solution is indeed
unique.

\section{Properties of the Solution}

In this section, we discuss various properties of the unique analytic space-time
$(M,g)$ found in the previous section which obeys the conditions of \S III with
$k=-1$ and $G(x)=-24m_0(x^2+4x)^{-5/2}$. Throughout this section, terms such as
`the solution', `the line element' etc. refer to this solution and its line
element etc. unless otherwise specified.

\subsection{Representation with an Elliptic Integral}
We show here that the function $e^\mu$ can be represented intrinsically by an 
elliptic integral; the basic results are from \cite{ksmh}.

Writing $\gamma=(x^2+4x)^{1/2}u$, we can obtain the following first integral
of (\ref{16a});
\begin{eqnarray} (x^2+4x)^2u_x^2&=&4(u^2-4m_0u^3) + A(t)\,,\label{5.1}\ee
where $A(t)$ is a function of integration.
Using (\ref{24}) above, we can compare powers of $x^{-1}$ in this last equation
to obtain
\[ A(t) = e^{-\beta(t)}\,.\]

A straightforward calculation shows that 
\[ (x^2+4x)u_x = -\frac{\gamma^2}{h}\frac{\partial R}{\partial r}\,,\]
where $R(r,t):=h(r)e^{\mu/2}$ is the radius of metric 2-spheres in the 
space-time, so to ensure $\partial R/\partial r >0$, 
we take the negative square root of (\ref{5.1}), which leads to an intrinsic elliptic
integral representation for $u$;
\[ \int \frac{du}{(4u^2-16m_0u^3+e^{-\beta})^{1/2}} = -\frac{1}{4}
\ln\left(\frac{x}{x+4}\right) +B(t)\,,\]
and $B(t)$ is the sole remaining function of integration. This term may involve $m_0$, and so cannot be determined from the other relevant limit, $m_0=0$ The integral on the left hand
side does not have a representation in terms of elementary functions for
$m_0\neq 0$, and so nor does our solution.

We note that it may be possible to determine the functions $A(t), B(t)$
in the elliptic integral which yield the correct asymptotic behaviour for $\mu$
{\it without} prior knowledge of the solution. The advantage is that we would not
need the existence proof of the previous section. However, this would be a rather difficult
problem involving inversion of asymptotic formulae for elliptic integrals. We feel
that the chosen method is the most direct, and has the advantage of dealing
with the general case, $G(x)=O(x^{-5})$.


\subsection{Energy Conditions}

Using the field equation (\ref{4}) and the 
first integral of the main equation (\ref{14}) as found in \S V-A above,
we can write the pressure (\ref{3}) as
\begin{eqnarray} 8\pi p&=&
-\frac{3}{4}\bedot^2+3e^{-\beta}-{\dot \mu}^{-1}\bedot(\btt+2e^{-\beta})
\,.\label{5.4}\ee
Notice that by the main theorem of \S IV, in the limit $\rin$, this coincides with
$8\pi p_0$, the pressure of the background RW universe. Thus all of the 
curvature tensor terms match up with those of the RW background in this limit.

Using (\ref{24}), we find that
\begin{eqnarray} 8\pi p&=&8\pi p_0 +(\btt + 2e^{-\beta})m_0e^{-\beta/2}e^{-3r} +
O(e^{-5r})\,.\label{5.5}\ee
Recall also that the energy density obeys $\rho(t)=\rho_0(t)$. We see then that
the question of whether or not $\rho$ and $p$ obey appropriate energy conditions
\cite{hawkell}
is, for sufficiently large values of $r$, equivalent to the same question
regarding $\rho_0$ and $p_0$.
Notice that if the weak energy condition is satisfied in the RW background,
then $\rho_0+p_0\geq0$, leading to $\btt+2e^{-\beta}\leq0$. 
Thus according to (\ref{5.5}), to first order, the presence of a central mass
in a $k=-1$ RW universe causes a {\it decrease} in the fluid pressure,
contrary to what one would expect.
This is in distinction to the situation in the $k=0$ model, where according to
(\ref{mcvp}) we can write
\[8\pi p=8\pi p_0 -\btt m_0e^{-\beta/2}r^{-1} + O(r^{-2})\,.\]
The weak energy condition in the RW background implies $\btt\leq0$, and so
the first order perturbation of the pressure is positive, as expected.
Indeed this behaviour is continued at all orders; $8\pi p\geq 8\pi p_0$
in the $k=0$ model, provided the weak energy condition holds in the RW background.

This latter situation is in line with what happens in the analogous situation
in Newtonian cosmology. The potential describing the physical scenario under
consideration is found (in this linear theory) by adding the potentials
$\phi_m = -mr^{-1}$ and $\phi_c=\rho(t)r^2/12$ for respectively a point 
particle of mass $m$ situated at $r=0$ and an isotropic cosmological model
with density $\rho(t)$ (see \cite{ellis} for the latter).
We take the potential to be
\[ \phi = \phi_m + \phi_c =-\frac{m}{r}+\frac{1}{12}\rho(t)r^2\,.\]
Then the pressure across the surface $\{S:r={\rm constant}\}$ at time $t$ is
\[ p(S)=\int_S -\vec{\bigtriangledown}{\phi}\cdot\vec{n}\,d_2S\,,\]
where $\vec{n}$ is the unit inward normal to $S$. This yields
\[ p(S) = 4\pi m+\frac{2}{3}\pi\rho r^3\,.\]
Thus we see that in Newtonian theory, the central mass makes a positive contribution
to the pressure.

The negative first order contribution in the $k=-1$ case could be cancelled
out by higher order terms, leaving a net positive contribution, but if not, 
it appears to be an interesting effect of the negative curvature of the
spatial sections of the RW background.

\subsection{Recovering the Schwarzschild Space-Time}

We have seen above how the line element of the RW background is recovered
by taking $m_0=0$. We show next how the line element of the exterior Schwarzschild
field arises in a natural way as a limiting case of our solution.

In McVittie's $k=0$ solution, the Schwarzschild field is obtained by setting 
${\dot{\beta}}$ equal to zero.
Then following a constant rescaling of the coordinate $r$, the line element
(\ref{mcv}) with $k=0$ is the isotropic form of the Schwarzschild line element
with mass parameter $m_0$. 
The procedure is quite natural; with $\bedot=0$, the energy density and pressure both
vanish, yielding a spherical vacuum which is by necessity, the exterior
Schwarzschild field. Note also that the expansion of the fluid flow lines
$\theta=\frac{3}{2}\bedot$ is then also equal to zero.

Carrying out the same procedure in the $k=-1$ case leads to
\[ \rho(t)=\rho_0(t) =\frac{3}{4}\bedot^2 - 3e^{-\beta}=0\,,\]
giving
\begin{eqnarray} e^\beta&=&(t+c)^2 \label{5.3} \ee
for some constant $c$.
Calculating $p_0(t)$, the pressure of the RW background, we find $p_0(t)=0$,
so that this space-time is (a portion of) Minkowski space-time.
Similarly, calculating $p(r,t)$ for our solution using this form of $\beta$ yields $p=0$,
and so the Ricci tensor vanishes. From (\ref{6}), the Weyl tensor remains non-zero,
and so by Birkhoff's theorem, $(M,g)$ is (a portion of) the exterior Schwarzschild
field. The Hawking mass of the Schwarzschild field is the Schwarzschild mass
parameter, and so $m_0$ in our solution is the Schwarzschild mass parameter.

Thus McVittie's solution for $k=0$, and our solution for the case $k=-1$, 
represents the Schwarzschild field embedded in a RW universe.

Another limiting case is of importance, namely when $\rho+p=0$, so that the
space-time is an Einstein space. In both cases ($k=-1,0$), we can explicitly verify that
$\rho+p=0$ implies that $p$ is constant. Again, the choice of $\beta(t)$ does not
affect the value of the Weyl tensor, and so by the `Birkhoff-with-a-cosmological-constant' 
theorem, space-time is a portion of the Schwarzschild--de Sitter cosmos. Thus
the solutions discussed here give genuinely cosmological (i.e. non-stationary) 
generalizations of this static space-time.

\subsection{Behaviour at Future Null Infinity}

As $\rin$ on the space-like hypersurfaces orthogonal to the fluid flow lines,
the line element of our solution approaches that of a RW space-time. The question
of how it behaves asymptotically along future null directions is more complicated,
but the following argument indicates that the space-time tends to a RW universe in
this limit. We show that the metric coefficients of our solution match up with
those of the RW background
as $\rin$ along future null directions
of the RW background, which are hence asymptotically future null 
directions of $(M,g)$.
We deal explicitly with the more complicated case $k=-1$; analogous results
hold for $k=0$. We consider first the description of $\scri^+$ in the
RW background. The following relies heavily on \cite{glass}.

Consider the RW background, whose line element may be written
\begin{eqnarray} ds^2&=&-dt^2+e^{\beta(t)}\left\{dr^2+\sinh^2rd\omega^2\right\}\nn\\
&=&\Omega^2(\eta)\left\{-d\eta^2+dr^2+\sinh^2rd\omega^2\right\}\,,\label{5.6}\ee
where $\Omega(\eta)=e^{\beta/2}(t)$,  $dt=\Omega(\eta)d\eta$
and $d\omega^2$ is the line element of the unit 2-sphere.
Define coordinates $u\,(0<u<\infty)$ and $\chi\,(0\leq \chi<\infty)$ by
\begin{eqnarray} u=e^{\eta-r}&\leftrightarrow&\eta=\frac{1}{2}\ln(u^2+2u\chi),\\
\chi=e^\eta\sinh r&\leftrightarrow &
r=\frac{1}{2}\ln(1+2\chi u^{-1}).\ee
In these coordinates, the line element (\ref{5.6}) assumes the form
\[ ds^2=F^2(u,\chi)\left\{-du^2-2dud\chi+\chi^2d\omega^2\right\}\,,\]
where
\[ F(u,\chi) = \Omega^2(\ln{(u^2+2u\chi)^{1/2}})(u^2+2u\chi)^{-1}\,.\]
Next, define $l=\chi^{-1}$ and introduce the non-physical line element
\begin{eqnarray} d\tilde{s}^2&=&H^2(u,l)ds^2\nn\\
&=& -l^2du^2+2dudl+d\omega^2\,,\label{5.7}\ee
where
\[ H(u,l)=l\Omega^{-1}(\ln{(u^2+2ul^{-1})^{1/2}})(u^2+2ul^{-1})^{1/2}\,.\]
Then, in the usual way, future null infinity of the RW space-time is identified
with the boundary $H=0$ of the space-time $(\tilde{V},\tilde{g})$ whose metric is given via the
line element (\ref{5.7}). If $H=0$ coincides with $l=0$, then a direct calculation
shows that $\scri^+$ is a shear-free null hypersurface. This depends upon $\Omega(\eta)$
being a sufficiently rapidly increasing function of its argument, which relates
to the conditions required of $e^{\beta}(t)$ to ensure convergence of the solution
in \S IV.
We note that for a perfect fluid with equation of state $p=\alpha\rho$, we have
\[ \Omega(\eta)=A\left( \sinh(\frac{3\alpha+1}{2}\eta)\right)^{2/(3\alpha+1)}\,,\]
for some constant $A$, which leads to
\begin{eqnarray} H(u,l) &=&A^{-1}2^{2/(3\alpha+1)}\\&&\times
\frac{l(u^2l+2u)}{\left( (u^2l+2u)^{(3\alpha+1)/2} - l^{(3\alpha+1)/2}\right)^{2/(3\alpha+1)}}\,,\ee
so that these conditions are satisfied if $3\alpha+1>0$.

This shows how to describe $\scri^+$ in the RW backgrounds for a large class of
such space-times. The importance of this for our situation is that it tells
us that as $\chi\rightarrow\infty$ along $u=$constant, we approach $\scri^+$ in
the RW universe.
To conclude this section, we simply note that in this limit, 
the metric coefficients of our solution 
approach those of the RW background, and all the curvature tensor terms
approach the background values. Thus our solution is asymptotically RW at
future null infinity; we already know it to be asymptotic to the RW background at
spacelike infinity.

\subsection{Singularities}
The solutions which we study here (McVittie's $k=0$ solution and our $k=-1$
version) have been shown to represent the Schwarzschild\, field embedded in a RW
universe. It is therefore natural to ask if these solutions have a central 
singularity and event horizon, and if so, how they are affected by the cosmic expansion.
We will treat this important issue in more depth elsewhere; we can give the following preliminary
results here.

For any space-time, the quantity $I=\Psi_0\Psi_4-4\Psi_1\Psi_3+3\Psi_2^2$ is
an invariant of the curvature. Here, we have for both $k=0,-1$
\begin{eqnarray} I&=&3\frac{m_0^2}{R^6}\,.\label{5.10}\ee
Thus we see that there is indeed an intrinsic curvature singularity at the centre
$R=0$. The coordinates we have used might not cover this region; this is immaterial
as (\ref{5.10}) derives from an invariant property of the curvature tensor, namely
$M_2(S)=m_0$ for all metric 2-spheres $S$.

Hayward \cite{hay} has shown that the Misner--Sharp gravitational energy is 
a useful tool for investigating singularities in spherical symmetry. One of the
equivalent definitions for this quantity is
\begin{eqnarray} E:=\frac{R}{2}(1-\chi)\,,\qquad \chi:=\nabla_aR\nabla^aR\,.\label{5.11}\ee
Carrying out a straightforward calculation which makes use of (\ref{4}) and the 
first integral (\ref{5.1}), we obtain the following nice results, which apply
to both $k=0$ and $k=-1$.
\begin{eqnarray} \chi&=&-\frac{8\pi}{3}R^2\rho(t)+1-2m_0R^{-1}\,,\label{5.12}\\
E&=&\frac{4\pi}{3}R^3\rho(t)+m_0\,.\label{5.13}\ee
These forms have the advantage of being coordinate independent; both $\rho$ and $R$
are invariantly defined quantities. (\ref{5.13}) is particularly satisfying;
the effect on the gravitational energy of the presence of a particle of mass
$m_0$ is an increase of exactly this amount.

Notice now that if $x$ is any point in the boundary $R=0$, then
\[ \lim_{\gamma\rightarrow x} E> 0\,,\]
along any curve $\gamma$ approaching $x$. Thus by a result of Hayward \cite{hay},
the central singularity is space-like and trapped, as in the Schwarzschild\, space-time.

In the case $k=0$, we see from (\ref{mcvp}) that there is also a curvature singularity
at $r=m/2$, which, intriguingly, corresponds to $R=2m_0$, the gravitational radius of the central mass. We see from (\ref{5.12}) that this is
a space-like hypersurface, and is surrounded by a trapped region.
The existence of this singularity is fundamentally different to the vacuum case,
and demands a thorough investigation of the singularity and horizon structure
of this space-time. These issues are currently being studied.

\subsection{Summary}
The solution we have found represents a point mass $m_0$ embedded in a $k=-1$
RW universe. When $m_0=0$, we obtain this RW background.
The energy-density is identical to that of the background, and the
zero-density limit gives Schwarzschild's space-time with mass
parameter $m_0$. The space-time is
asymptotic to the RW universe at infinity and contains a space-like
singularity at the centre.

\section{Comments}

We have given a prescription above for embedding the Schwarzschild field in a 
give open RW universe. Consider the converse problem. Given a spherically
symmetric shear-free perfect fluid space-time $(V,g)$, how do we
know if $(V,g)$ represents a point mass in a RW universe and if it does, how do
we identify that RW universe? This presents us with a gauge problem.
``Suppose we consider the lumpy universe model $S$, not knowing how the [background]
model $\bar{S}$ was used to make the construction; can we uniquely recover $\bar{S}$
from $S$?''\cite{ellis-bruni}. In fact the answer to this question is yes. Calculate the
Hawking mass for an arbitrary metric 2-sphere of $(V,g)$. If the result is not
a constant, then $(V,g)$ does not represent a point mass in a RW universe.
If the result is a constant ($m_0$ say), and if further when $m_0=0$, $(V,g)$ is
a RW universe, we may proceed. This solves the gauge problem by identifying
the background model. It remains then to check if $(V,g)$ satisfies the remaining
parts of conditions ${\bf C2, C3}$ with respect to this well defined RW
background.

In a previous paper \cite{mcvsource}, we interpreted certain space-times
as being extended sources for the McVittie field in the three cases $k=0,\pm 1$.
To further
investigate the occurence of singularities and horizons in $(M,g)$, it would be interesting
to determine if a collapsing fluid can be used as a source. This would allow us to
interpret $(M,g)$ as the end state of the spherical collapse of a massive body
in an expanding universe and may throw some light on the issue of what, if any, the effect
of this expansion is on the collapse. 
This still leaves the problem of whether the space-time is of black-hole 
(collapsed object surrounded by an event horizon) or white-hole (lagging 
core of an expanding universe) type. This issue is to be addressed in a subsequent 
paper in which the horizon, singularity and asymptotic structure of these 
space-times is analysed. 

The nature of the solution we have found has opened up these interesting questions.
However, our main purpose was to give a clear physical interpretation of some solutions
of Einstein's equation. In particular, we hope to have given such
for McVittie's solution, which in the case $k=0$ does indeed represent a point 
mass in an RW universe; some authors have contested this interpretation
\cite{its,suss3}. We have seen how Hawking's mass was a useful tool
in this. Our aim now is to use this tool in an attempt to identify solutions
representing the embedding of other objects (cosmic strings, the Reissner-Nordstrom and
Kerr fields) in RW universes. We note that some of the solutions \cite{othersols}
given previously
which it was claimed represent such do not reproduce McVittie's solution 
in the
$k=0$ case or our solution in the $k=-1$ case in the appropriate limit 
(charge-free and non-rotating). This may be an inherent discontinuous feature 
of solutions of the field equations in such situations. However it leads to the
suspicion that these solutions do not satisfy a set of conditions analogous to 
${\bf C1-C3}$ which clearly determine their physical interpretation.

\section*{Acknowledgement}
I am indebted to Chris Luke for his invaluable assistance with the proof in \S IV.



\begin{thebibliography}{tails}


\bibitem[1]{israel}
W. Israel, Nuovo Cimento {\bf 44}, 1 (1966).

\bibitem[2]{sato}
H. Sato, in {\it General Relativity and Gravitation} ed. B. Bertotti, F. de 
Felice and A. Pascolini (Reidel, Boston, 1984).

\bibitem[3]{vilenkin}
A. Vilenkin and E.~P.~S. Shellard,
{\it Cosmic Strings and Other Topological Defects}
(Cambridge University Press, Cambridge, 1994).

\bibitem[4]{mcv}
G.~C.~ McVittie, Mon. Not. Roy. Astron. Soc. {\bf 93}, 325 (1933).

\bibitem[5]{hawk}
S.~W. Hawking, J. Math. Phys. {\bf 9}, 598 (1968).

\bibitem[6]{othersols}
P.~C. Vaidya and Y.~P. Shah, Current Sci. (India) {\bf 36}, 120 (1967).
P.~C. Vaidya, Pramana {\bf 8}, 512 (1977).
L.~K. Patel and S.~S. Koppar, Acta Phys. Hung. {\bf 64},
353 (1988)

\bibitem[7]{kraz}
A. Krasinski,
{\it Inhomogeneous Cosmological Models}
(Cambridge Universtiy Press, Cambridge, 1997).

\bibitem[8]{ellis-bruni}
G.~F.~R. Ellis and M. Bruni, Phys. Rev. {\bf D 40}, 1804 (1989).

\bibitem[9]{e-s}
A. Einstein and E.~G. Straus, Rev. Mod. Phys. {\bf 17}, 120 (1945);
{\bf 18}, 148 (1946).

\bibitem[10]{seno}
J.~M.~M. Senovilla and R. Vera, Phys. Rev. Lett. {\bf 78}, 2284 (1997).

\bibitem[11]{mars}
M. Mars, ``Axially symmetric Einstein--Straus models'', Phys. Rev. {\bf D 57} (1998).

\bibitem[12]{ksmh}
D. Kramer, H. Stephani, M.~A.~H. MacCallum and E. Herlt,
{\it Exact Solutions of Einstein's Field Equations}
(Cambridge University Press, Cambridge, 1980).

\bibitem[13]{hog1}
P.~A. Hogan, Astroph. J. {\bf 360}, 315 (1990).

\bibitem[14]{ray}
A.~K. Raychaudhuri, {\it Theoretical Cosmology}
(Clarendon, Oxford, 1979).

\bibitem[15]{bondi}
H. Bondi, M.~G.~J. van der Burg  and A.~W.~K. Metzner,
Proc. Roy. Soc. London A{\bf 269}, 21 (1962).

\bibitem[16]{haymass}
S.~A. Hayward, Phys. Rev. {\bf D 49}, 831 (1994).

\bibitem[17]{remmert}
R. Remmert, {\it Theory of Complex Functions}
Graduate Texts in Mathematics 122
(Springer-Verlag, New York, 1991).

\bibitem[18]{apostol}
T.~M. Apostol, {\it Mathematical Analysis}
(Addison-Wesley, Reading, 1974).

\bibitem[19]{hawkell}
S.~W. Hawking and G.~F.~R. Ellis,
{The Large Scale Structure of Space-Time}
(Cambridge University Press, Cambridge, 1973).

\bibitem[20]{ellis}
G.~F.~R. Ellis, in {\it General Relativity and Cosmology},
proceedings of the XLVII Enrico Fermi Summer School, ed. R.~K. Sachs
(Academic Press, New York, 1971).

\bibitem[21]{glass}
E.~N. Glass, J. Math. Phys. {\bf 12}, 1222 (1971).

\bibitem[22]{hay}
S.~A. Hayward, Phys. Rev. {\bf D 53}, 1938 (1996).

\bibitem[23]{mcvsource}
B. Nolan, J. Math. Phys. {\bf 34}, 178 (1993).

\bibitem[24]{its}
M. Ferraris, M. Francaviglia and A. Spallicci,
Nuovo Cim. B {\bf 111}, 1031 (1996).

\bibitem[25]{suss3}
R.~A. Sussmann, J. Math. Phys. {\bf 29}, 1177 (1988).
\end{thebibliography}
\end{document}